\documentclass[pre,10pt,twocolumn,superscriptaddress,showpacs,showkeys]{revtex4-1}
\usepackage{amsmath}
\usepackage{latexsym}
\usepackage{amssymb}
\usepackage{lipsum, babel}
\usepackage{mathtools}
\usepackage{braket}
\usepackage{graphicx}
\usepackage[caption=false]{subfig}
\usepackage{wrapfig}
\usepackage{dsfont}
\usepackage{etoolbox}
\usepackage[colorlinks=true, citecolor=blue, urlcolor=blue]{hyperref}
\usepackage{float}
\usepackage{amsfonts}
\usepackage{soul}
\usepackage[T1]{fontenc}

\begin{document}
\title{Bond current in a mesoscopic ring -- signature of decoherence due to classical and quantum noise}
\author{Sushanta Dattagupta}
\email{sushantad@gmail.com}
\affiliation{Bose Institute, Kolkata 700054, India}
\author{Tanmay Saha}
\email{sahatanmay@imsc.res.in}
\affiliation{Optics and Quantum Information Group, The Institute of Mathematical Sciences, CIT Campus, Taramani,
Chennai 600113, India}
\affiliation{Homi Bhabha National Institute, Training School Complex, Anushakti Nagar, Mumbai 400085, India}

\begin{abstract}
A three-site mesoscopic ring provides an ideal setting for an exact calculation of the bond current when the ring is threaded by an Aharonov-Bohm flux. The bond current is a measurable outcome of the coherent properties of the quantum phase. However the coherence is impeded by noise when the ring is put in contact with an environment. This coherence-to-incoherence transition is analyzed in detail here for both classical (Gaussian and telegraphic) and quantum noise and a comparative assessment is made when the quantum noise is governed by a spin-boson Hamiltonian of dissipative quantum mechanics.
\end{abstract}

\keywords{Aharonov-Bohm phase, mesoscopic bond current, Gaussian and telegraphic processes, quantum noise}

\maketitle

\section{Introduction}
\label{introduction}
 Mesoscopic devices are systems in which basic quantum mechanical principles can be tested in the laboratory \cite{alma991051260949706011,alma9926223099805776}. Besides, they can be put to useful applications toward the processing of quantum information. While the latter hinges on the coherent properties of the quantum phase, the flip side is that mesoscopic systems being small are inevitably in coupling with their environment which is in effect noisy. The noise at elevated temperatures can be described by classical stochastic processes but, and more importantly for low-temperature phenomena, needs to be tackled quantum mechanically. A theoretical understanding of what the noise does to the otherwise pure quantum dynamics and how it can be effectively controlled are important in ensuring the efficacy of such mesoscopic devices.

\begin{figure}
\begin{center}
\includegraphics[width=0.4\textwidth]{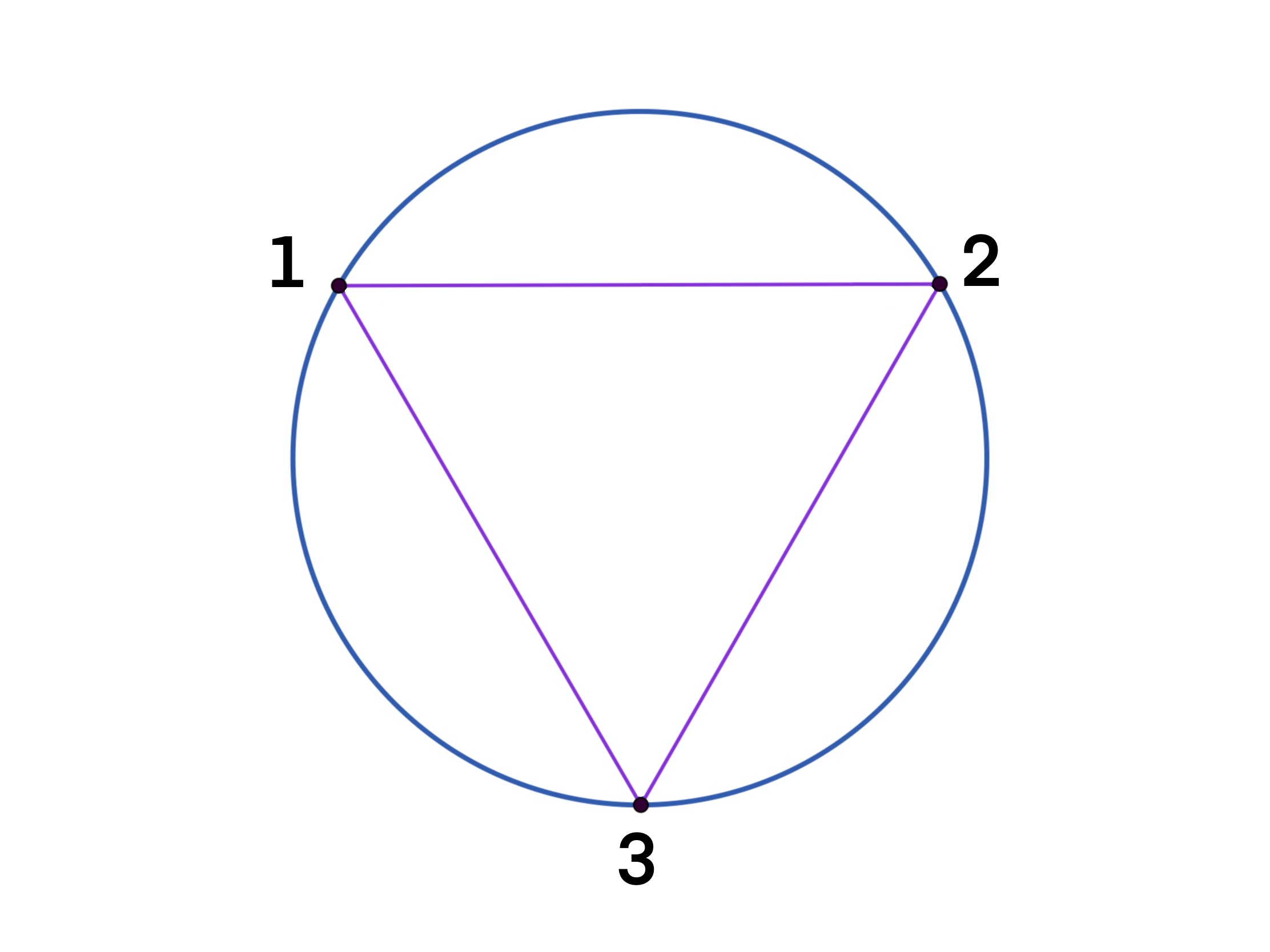}
\caption{(color online) Schematic of a three-site mesoscopic ring.
}
\label{Ring}
\end{center}
\end{figure}

In this paper we study this interplay of quantum coherence and noise-induced decoherence \cite{PhysRevB.67.195320, 2006, 2009,  2012}. However, in order to keep the discussion at an analytically tractable form, we consider a model system of a three-site ring which is characterized by the interesting occurrence of a ‘persistent’ (without a battery) bond current when the ring is threaded by the Aharonov-Bohm flux due to a perpendicular magnetic field \cite{AandB1959, SDG2018}. The Aharonov-Bohm effect is a distinct imprint of the quantum phase and the concomitant presence of coherence \cite{Tonomura1998TheQW}. After the surprising effect was theoretically predicted in $1959$ the resultant bond current has been measured in the laboratory \cite{PhysRevLett.64.2074, PhysRevLett.67.3578,PhysRevLett.70.2020, PhysRevLett.86.1594}. The bond current, a transport property is intriguingly related to orbital magnetism, an attribute of thermodynamics \cite{PhysRevB.51.11584}. To keep the analysis simple the ring is assumed to be symmetric, i.e., all site energies are the same (as would happen in the absence of any disorder) in which the dynamics takes place through quantum tunneling of a single electron from site to site (Fig. \ref{Ring}).
As far as the environmental influence is concerned we adopt the well-known classical stochastic processes of Gaussian and Telegraph forms \cite{agarwal1983stochastic} and for quantum noise, the spin-boson model of dissipative quantum systems \cite{doi:10.1142/8334, DattaguptaPuri,PhysRevB.78.201302,2016}.

With the known result for the bond current, previewed in the beginning of the next section, our aim in this paper is to endow the ring Hamiltonian with explicitly time-dependent properties in order to account for the fluctuating effect of the environment, and calculate the time-dependent current. Thus, the tunneling energy $J$ is either written in terms of a stochastic process $f(t)$ as $[J + \Delta f(t)]$, or the variable $f(t)$ is viewed to live in a large Hilbert space of a quantum heat bath representing the fluctuating influence of the environment of the ring.

Given these introductory remarks on the scope of the present investigation, the paper is organized as follows. The basic formulation needed for fulfilling the objective of this study is presented in Sec. \ref{formulation}. In Sec. \ref{gaussian}, we treat the case of classical noise for the well-known situation in which $f(t)$ is a Gaussian stochastic process and adapt it for the computation of the bond current that allows us also to introduce the cumulant expansion scheme for the stochastically averaged time-evolution operator  \cite{10002987861,doi:10.1143/JPSJ.17.1100,doi:10.1063/1.1703941,dattagupta1987relaxation}. As contrasting examples, Sec. \ref{STP} and \ref{ASTP} are devoted to a two-state Telegraph process (TP) (symmetric and asymmetric respectively) description of $f(t)$ and once again, certain exact results are presented for graphical comparison with the Gaussian case of Sec. \ref{gaussian}. In Sec. \ref{quantum}, we move on to the quantum domain and view $f(t)$ in the language of a spin-boson system. The results for classical noise, derived in Sec. \ref{ClassicalNoise}, and those for the quantum noise, dealt with in Sec. \ref{quantum}, are now analysed together. Also the conditions under which a TP emerges from a spin-boson Hamiltonian are elucidated in this section. Finally, Sec. \ref{conclusions} contains our concluding remarks.
\section{Formulation}
\label{formulation}
The Hamiltonian for a symmetric three-site ring can be written as 
\begin{equation}
\boldsymbol{\mathcal{H}} = -J(\ket{1}\bra{2}+\ket{2}\bra{3}+\ket{3}\bra{1}+h.a.),\label{intro1}
\end{equation}
where $h.a.$ denotes Hermitian adjoint, $\ket{i}$ denotes the state of the site $i$ for $i=1,2,3$, and $J$ is a (real) tunneling energy, assumed the same between sites, which however becomes complex in the presence of a perpendicular magnetic field when the Hamiltonian becomes
\begin{equation}
\boldsymbol{\mathcal{H}_{R}} = -J[\exp{(-i\phi)} (\ket{1}\bra{2}+\ket{2}\bra{3}+\ket{3}\bra{1})+h.a.],\label{intro2}
\end{equation}
where the subscript $R$ stands for ‘ring’ while $\phi$ is the Aharonov-Bohm phase (assumed identical for a pair of sites) given in terms of the charge $e$ of the electron, the speed of light $c$, the magnetic field $B$ and the area $A$ of the triangle shown in Fig. \ref{Ring} as
\begin{equation}
\phi = \frac{eBA}{c}.\label{intro3}
\end{equation}
(Here the Planck constant has been set equal to unity.)

The rate of leakage of the electron from any given site, say $1$, is given by
\begin{align}
\frac{d}{dt}\ket{1}\bra{1} = i[\boldsymbol{\mathcal{H}}, \ket{1}\bra{1}] = &-iJ(e^{i\phi}\ket{2}\bra{1}-e^{-i\phi}\ket{1}\bra{2})\nonumber\\
 &- iJ(e^{-i\phi}\ket{3}\bra{1}-e^{i\phi}\ket{1}\bra{3}).\label{intro4}
\end{align}
The first term on the right of Eq. (\ref{intro4}) within the round brackets has the interpretation of the current from site $1$ to site $2$ (clockwise) while the second term the current from the site $1$ to $3$ (counter-clockwise). Since the particle number is conserved the left-hand side in the ground quantum mechanical state must vanish. Thus, the expectation value of the electric current, say from site $1$ to $2$ – which can be measured by sticking-in an ammeter between the sites $1$ and $2$  – is given by
\begin{align}
I &= 2eJ\boldsymbol{\langle} i(e^{i\phi}\ket{2}\bra{1}-e^{-i\phi}\ket{1}\bra{2})\boldsymbol{\rangle}\nonumber\\
  & = 4eJ~Im\boldsymbol{\langle} e^{i\phi}\ket{2}\bra{1}\boldsymbol{\rangle},\label{intro5} 
\end{align}
where $e$ is the electronic charge. The angular brackets $\boldsymbol{\langle...\rangle}$ denote the expectation value in the ground state of the Hamiltonian in Eq. (\ref{intro2}). The latter being a $3\times 3$ matrix can be easily diagonalized to yield the eigenvalues and eigenfunctions over which the expectation value can be computed \cite{SDG2018}. Hence,
\begin{align}
&I = 4eJ~Im~\{e^{i\phi}\boldsymbol{\langle}\boldsymbol{j}_{12}\boldsymbol{\rangle}\},~~~~~\textnormal{here,}\label{intro6}\\
& \boldsymbol{j}_{12} = \ket{2}\bra{1}, \boldsymbol{\langle} \boldsymbol{j}_{12}\boldsymbol{\rangle} = \sum_{m} \braket{m|\boldsymbol{j}_{12}|m},\label{intro7}
\end{align}
where the summation is over the different eigenfunctions $m$ of $\boldsymbol{\mathcal{H}_{R}}$ (three, in this case) that can be written down as \cite{SDG2018}
\begin{align}
\ket{m = +} &=\frac{1}{2\sqrt{3}}\begin{pmatrix}
						-1-i\sqrt{3}\\
						-1+i\sqrt{3}\\
						2\\
						\end{pmatrix},\nonumber\\
\ket{m= 0} &=\frac{1}{\sqrt{3}}\begin{pmatrix}
						1\\
						1\\
						1\\
						\end{pmatrix},\nonumber\\
\ket{m = -} &=\frac{1}{2\sqrt{3}}\begin{pmatrix}
						-1+i\sqrt{3}\\
						-1-i\sqrt{3}\\
						2\\
						\end{pmatrix},\label{intro8}
\end{align}
associated with the eigenvalues
\begin{align}
\lambda_{+} &= J(\cos{\phi}+\sqrt{3}\sin{\phi}),\nonumber\\
\lambda_{0} &= -2J\cos{\phi},\nonumber\\
\lambda_{-} &= J(\cos{\phi}-\sqrt{3}\sin{\phi}).\label{intro9}
\end{align}
Some straightforward algebra yields
\begin{equation}
I = 4eJ\sin{\phi}.\label{intro10}
\end{equation}
From $\phi(B) = -\phi(-B)$ follows that $I(B) = -I(-B)$, explicitly demonstrating the effect of time-reversal.

In the presence of a stochastic process $f(t)$ the Hamiltonian can be modified as
\begin{align}
&\boldsymbol{\mathcal{H}_{R}}(t) = -[J+\Delta f(t)]\boldsymbol{\it{ \Pi}},\label{intro11}\\ &\textnormal{where,}\nonumber\\
&\boldsymbol{\it{\Pi}} = [\exp{(-i\phi)} (\ket{1}\bra{2}+\ket{2}\bra{3}+\ket{3}\bra{1})+h.a.].\label{intro12}
\end{align}
It is evident that $[\boldsymbol{\mathcal{H}_{R}}(t),\boldsymbol{\mathcal{H}_{R}}(t')] = 0$, which implies that $\boldsymbol{\mathcal{H}_{R}}(t)$, once diagonalized, will remain diagonal for all times.  The physical meaning is there is no energy exchange between the system at hand and its surrounding bath giving rise to the so-called “adiabatic decoherence” \cite{1998, Gangopadhyay_2001, book:909268}.

From Eq. (\ref{intro6}) and (\ref{intro7}) the time-dependent current can be computed from
\begin{align}
&\boldsymbol{\langle}\boldsymbol{j}_{12}(t)\boldsymbol{\rangle} = \sum_{m} \bra{m}(e^{i\int_{0}^{t}dt'\boldsymbol{\mathcal{H}_{R}}(t')}\boldsymbol{j}_{12}e^{-i\int_{0}^{t}dt'\boldsymbol{\mathcal{H}_{R}}(t')})_{av}\ket{m}\label{intro13}\\
&\textnormal{while}~~~\boldsymbol{\langle}\boldsymbol{j}_{12}(0)\boldsymbol{\rangle} = \sum_{m}\braket{m|2\rangle\langle1|m},\label{intro14}
\end{align}
where $(…)_{av}$ denotes either a stochastic average (for classical noise) or a quantum bath-average (for quantum noise), specified in Sec. \ref{quantum}. Because $\boldsymbol{\mathcal{H}_{R}}(t)$ can be diagonalized once and for all, we have
\begin{equation}
\boldsymbol{S}^{\dagger}\boldsymbol{\mathcal{H}_{R}}(t)\boldsymbol{S} = -[J+\Delta f(t)]\Tilde{\boldsymbol{\it{\Pi}}},\label{intro15}
\end{equation}
where, $\boldsymbol{S}=\frac{1}{2\sqrt{3}}\begin{pmatrix}
					    -1-i\sqrt{3} & 2 & -1+i\sqrt{3}\\
					    -1+i\sqrt{3} & 2 & -1-i\sqrt{3}\\
					    2                 & 2 & 2\\
					    \end{pmatrix}$ and $\Tilde{\boldsymbol{\it{\Pi}}}=diag\big(-(\cos{\phi}+\sqrt{3}\sin{\phi}),~2\cos{\phi},~-(\cos{\phi}-\sqrt{3}\sin{\phi})\big)\equiv diag\big(\Tilde{\lambda}_{+},~\Tilde{\lambda}_{0},~\Tilde{\lambda}_{-}\big).$

From Eq. (\ref{intro13}),
\begin{align}
\boldsymbol{\langle}\boldsymbol{j}_{12}(t)\boldsymbol{\rangle}~~~~~~~~&\nonumber\\
= \sum_{mnn'}\boldsymbol{S}_{mn}\Big(&\exp\Big\{i\Big[Jt+\Delta\int_{0}^{t}dt' f(t')\Big]\Big(\frac{\lambda_{nn'}}{J}\Big)\Big\}\Big)_{av}\nonumber\\
&\bra{n}\boldsymbol{S}^{\dagger}\boldsymbol{j}_{12}\boldsymbol{S}\ket{n'}\boldsymbol{S}^{\dagger}_{n'm}, ~~\lambda_{nn'}=\lambda_{n}-\lambda_{n'}.\label{intro16}
\end{align}
Using the site representation
\begin{equation}
\ket{1}=\begin{pmatrix}
	     1\\
	     0\\
	     0\\
	     \end{pmatrix},~\ket{2}=\begin{pmatrix}
	     					0\\
	     					1\\
	    					0\\
	    					\end{pmatrix},~\ket{3}=\begin{pmatrix}
										    0\\
										    0\\
										    1\\
										    \end{pmatrix}.\label{intro17}
\end{equation}
and some algebra, we can show that
\begin{align}
&\bra{n}\boldsymbol{S}^{\dagger}\boldsymbol{j}_{12}\boldsymbol{S}\ket{n'} = \frac{-1+i\sqrt{3}}{2}\delta_{n+}\delta_{n'-},~\textnormal{and}\nonumber\\
&\sum_{m}\boldsymbol{S}_{m+}\boldsymbol{S}_{-m}\bra{+}\boldsymbol{S}^{\dagger}\boldsymbol{j}_{12}\boldsymbol{S}\ket{-} = 1.\label{intro18}
\end{align}
From Eq. (\ref{intro6}) and (\ref{intro16}) then, the time-dependent current can be written as 
\begin{align}
&I(t) = 4eJ~Im\{e^{i\phi}\exp(i\lambda_{+-}t)[\boldsymbol{U}(t)]_{av}\},~~\textnormal{where,}\label{intro19}\\
&[\boldsymbol{U}(t)]_{av} = \Big[\exp\Big(i\Delta\int_{0}^{t}dt' f(t')\Big(\frac{\lambda_{+-}}{J}\Big)\Big)\Big]_{av},\label{intro20}\\
&\textnormal{and,}~~\lambda_{+-}= 2\sqrt{3}J\sin{\phi}.\nonumber
\end{align}
Clearly, at time $t = 0$, the current matches with the initial value expression given in Eq. (\ref{intro10}). Here $[\boldsymbol{U}(t)]_{av}$ is the averaged time-evolution operator the calculation of which will be the focus of attention in the subsequent sections.

A remark is in order concerning the physical meaning of the initial value $I(0)$ and the final value $I(\infty)$ on the current. Recalling that the Aharonov-Bohm phase originates from a magnetic flux, the latter is envisaged to have been switched-on, at an infinite past. Thus, at time $t=0$, the ring is expected to come to a steady state (and we do not enquire how) wherein usual quantum mechanics applies, yielding the bond current $I(0)$ given by Eq. (\ref{intro10}). Next, the equilibrium is disturbed by switching-on, at $t=0$, a stochastic process $f(t)$ giving rise to the operative Hamiltonian in Eq. (\ref{intro11}). The further time evolution has two parts: $(a)$ a ‘coherent’ one dictated by the bare $J$ in Eq. (\ref{intro11}), and $(b)$ an ‘incoherent’ one governed by $f(t)$. Hence, in the infinite future, the disruptive effects of $f(t)$ are expected to lead to a zero current: $I(\infty)=0$, as is borne out by numerical plots given below.
\section{Bond current under a classical noise}
\label{ClassicalNoise}
\subsection{Gaussian case}
\label{gaussian}
As stated at the outset our first task is to assume that the environment of the mesoscopic ring is a classical heat bath that creates a fluctuating tunneling energy governed by a Gaussian stochastic process. For calculating the averaged time-evolution operator it is most convenient then to invoke the cumulant expansion theorem \cite{agarwal1983stochastic,10002987861,SDGdiffusion} because, for a Gaussian process, all cumulants beyond the second one are zero. The cumulant expansion theorem will turn out to be handy for the later treatment of quantum noise, in Sec. \ref{quantum} as well. In Eq. (\ref{intro20}) then
\begin{align}
[\boldsymbol{U}(t)]_{av} = \exp\{i\delta &\int_{0}^{t}dt' \boldsymbol{\langle\langle} f(t')\boldsymbol{\rangle\rangle}\nonumber\\
-\delta^{2}&\int_{0}^{t}dt'\int_{0}^{t'}dt''\boldsymbol{\langle\langle} f(t')f(t'')\boldsymbol{\rangle\rangle}\},\label{gaussian1}
\end{align}
where the double angular brackets $\boldsymbol{\langle\langle...\rangle\rangle}$ denote cumulant averages. Because the mean noise vanishes the first integral in the exponent is zero while the Doob's theorem yields \cite{SDGdiffusion}
\begin{equation}
\boldsymbol{\langle\langle} f(t')f(t'')\boldsymbol{\rangle\rangle} = \boldsymbol{\langle} f^{2}\boldsymbol{\rangle} \exp (-\gamma t),\label{gaussian2}
\end{equation}
$\gamma$ being the dissipation parameter, and
\begin{equation}
\delta \equiv 2\sqrt{3}\Delta \sin{\phi}.\label{gaussian3}
\end{equation}
Carrying out the double integrals \cite{PhysRevB.101.184308} in Eq. (\ref{gaussian1}) and substituting in Eq. (\ref{intro19}) the current is given by
\begin{equation}
I(t) = 4eJ \sin \Big[\phi +\frac{\delta Jt}{\Delta}\Big] e^{-\frac{\delta^{2}}{\gamma^{2}}(\gamma t-1+e^{-\gamma t})}.\label{gaussian4}
\end{equation}

This equation allows for separate discussion of \textit{weak} and \textit{strong dissipation}. In the \textit{weak damping} case $(\gamma t<<1),$
\begin{equation}
I(t) = 4eJ \sin \Big[\phi +\frac{\delta Jt}{\Delta}\Big] e^{-\frac{\delta^{2}t^{2}}{2}(1-\frac{\gamma t}{3})},\label{gaussian5}
\end{equation}

\begin{figure*}[]
     \centering
     \subfloat[]{\includegraphics[width=0.50\textwidth]{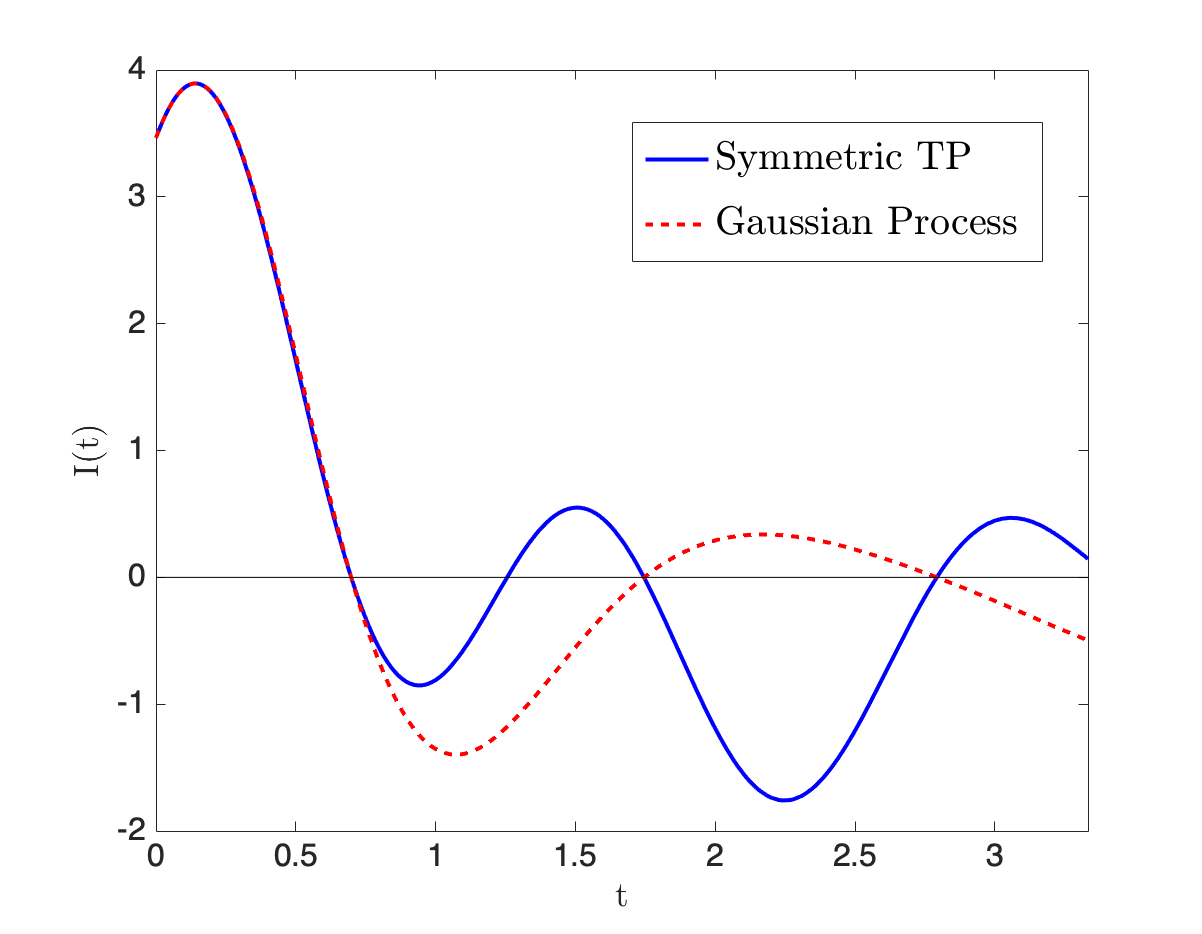}\label{currentweak}}
     \subfloat[]{\includegraphics[width=0.52\textwidth]{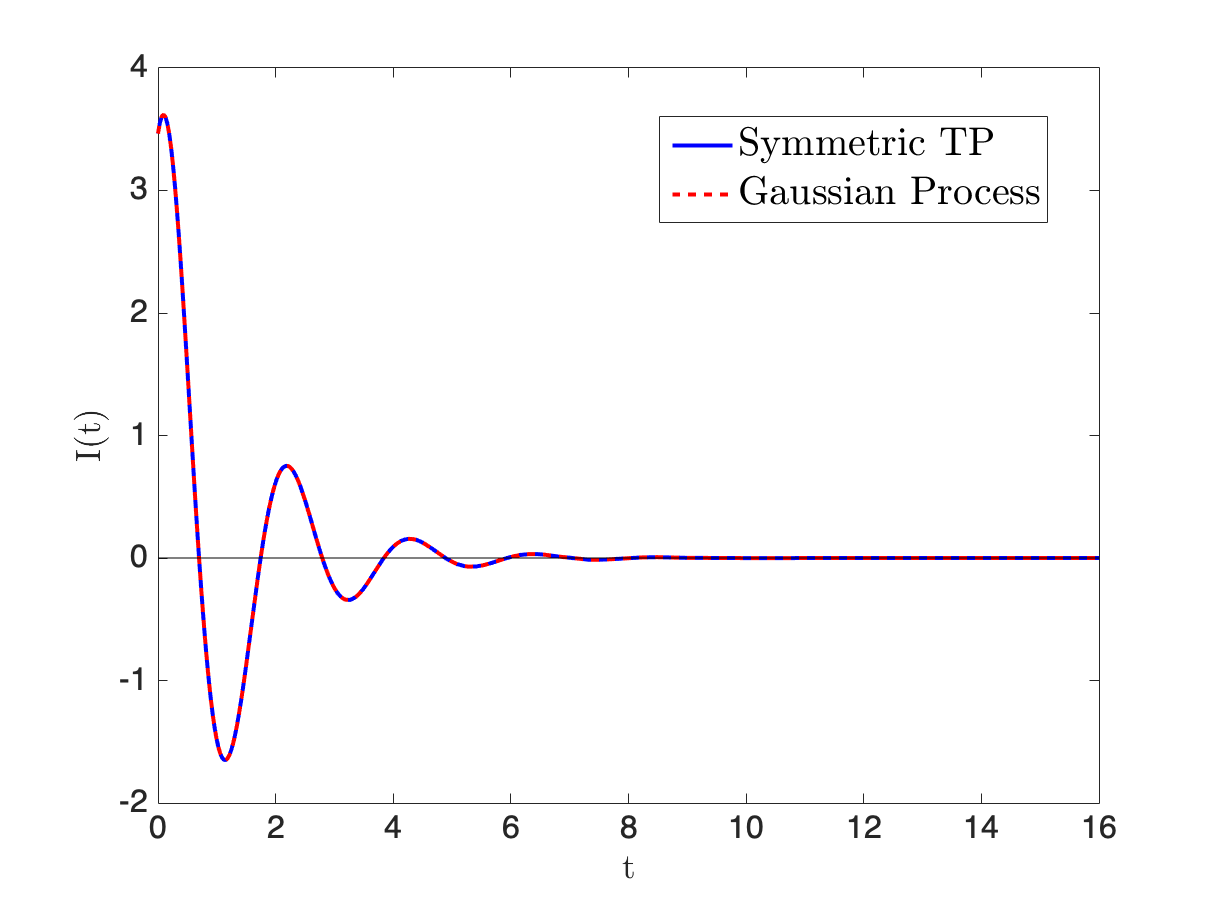}\label{currentstrong}}
     \caption{(color online) Plot of Persistent current with time for a comparative study between Symmetric TP (solid blue line) and Gaussian Process (dashed red line): (a) for \textit{weak dissipation}, the respective parameters are $2\nu=\gamma=\frac{3}{4}$ (b) for \textit{strong dissipation}, the parameters are  $2\nu=\gamma= 3.$ For both of the plots, $e=1=J, \Delta = \frac{1}{2}, \phi = \frac{\pi}{3},$ and $\delta$ is given in Eq. (\ref{gaussian3}).}\label{current}
\end{figure*}
  
whereas \textit{strong damping} $(\gamma t>>1)$ yields
\begin{equation}
I(t) = 4eJ \sin \Big[\phi +\frac{\delta Jt}{\Delta}\Big] e^{-\frac{\delta^{2}}{\gamma}t}.\label{gaussian6}
\end{equation}

In the next subsection we will compare these results with those for the symmetric TP.\\
\subsection{Telegraphic process -- symmetric case}
\label{STP}
As shown by Blume \cite{PhysRev.174.351} the telegraph process (TP) is a special case of a stationary Markov process in which the stochasticity is restricted to two states $|+)$ and $|-)$ (denoted by $|a)$, $|b)$...below), i.e., $f(t)$ jumps between two values $+1$ and $-1$. The Laplace transform of the averaged time-development operator in Eq. (\ref{intro20}) is given by
\begin{equation}
[\boldsymbol{U}(s)]_{av}=\sum_{ab}p_{a}(a|(s\mathds{1}-i\boldsymbol{F}-\boldsymbol{W})^{-1}|b),\label{stp01}
\end{equation}
where $s$ is the Laplace transform variable, $p_{a}$ is the \textit{a-priori} probability of the stochastic state $|a)$, $\boldsymbol{F}$ is a $2\times2$ diagonal matrix with elements given by the two allowed frequencies:
\begin{equation}
\boldsymbol{F}=\begin{pmatrix}
			\delta   &   0\\
			0          &   -\delta\\
			\end{pmatrix}.\label{stp02}
\end{equation}
The central quantity for the underlying Markov process is the rate matrix $\boldsymbol{W}$ given by
\begin{equation}
\boldsymbol{W}=2\nu\begin{pmatrix}
				-p_{-} & p_{-}\\
				p_{+} & -p_{+}\\
				\end{pmatrix},\label{stp03}
\end{equation}
where $\nu$ is the jump rate. The matrix of $\boldsymbol{W}$ is constructed on the basis of two distinct attributes:
\begin{align}
(i)~~\textnormal{conservation of probability}~~~~~&\nonumber\\
\sum_{b}(a|\boldsymbol{W}|b) =&0,~~\textnormal{all}~a’s,\label{stp04}\\
\textnormal{and}~(ii)~~\textnormal{detailed balance of transitions}&\nonumber\\
p_{a}(a|\boldsymbol{W}|b) =&p_{b}(b|\boldsymbol{W}|a).\label{stp05}
\end{align}
The jump matrix for a TP has the interesting property \cite{dattagupta1987relaxation}:
\begin{equation}
\boldsymbol{W} = 2\nu(\boldsymbol{T}-\mathds{1}),\label{stp06}
\end{equation}
where the transition matrix $\boldsymbol{T}$ satisfies
\begin{equation}
(a|\boldsymbol{T}|b) = p_{b},\label{stp07}
\end{equation}
independent of the initial state $|a)$. This special property of the $\boldsymbol{T}$- matrix allows  $[\boldsymbol{U}(s)]_{av}$ to be expressed in terms of the ‘static’ time-development operator \textit{albeit} with a self-energy correction that has the structure of a random phase approximation\cite{dattagupta1987relaxation}:
\begin{equation}
[\boldsymbol{U}(s)]_{av} = \frac{[\boldsymbol{U}_{0}(s+2\nu)]_{av}}{1-2\nu[\boldsymbol{U}_{0}(s+2\nu)]_{av}},\label{stp08}
\end{equation}
where,
\begin{equation}
[\boldsymbol{U}_{0}(s+2\nu)]_{av} = \sum_{a} p_{a}(s+2\nu-i\boldsymbol{F}_{a})^{-1}.\label{stp09}
\end{equation}
We are now ready to discuss the symmetric case in which the \textit{a-priori} probabilities of the occurrence of the two assigned values of $f(t)$ are identical, i.e., $p_{+} = p_{-} = \frac{1}{2}.$ In that case Eq. (\ref{stp08}) and (\ref{stp09}) yield, after some algebra
\begin{equation}
[\boldsymbol{U}(s)]_{av} = \frac{(s+2\nu)}{s(s+2\nu)+\delta^{2}},\label{stp10}
\end{equation}
from which the corresponding expression in the time-domain can be easily written down:
\begin{align}
[\boldsymbol{U}(t)]_{av} = \frac{1}{2}&\Big\{1-\frac{\nu}{\sqrt{\nu^2 -\delta^2}}\Big\}e^{-\nu t-\sqrt{\nu^2 -\delta^2}t}\nonumber\\
+\frac{1}{2}&\Big\{1+\frac{\nu}{\sqrt{\nu^2 -\delta^2}}\Big\}e^{-\nu t+\sqrt{\nu^2 -\delta^2}t}.\label{stp11}
\end{align}
It may be pertinent to mention here that $[\boldsymbol{U}(t)]_{av}$, directly obtainable from Eq. (\ref{stp01}) by Laplace-inverting, can be straightaway calculated in the time domain, in order to arrive at Eq. (\ref{stp11}), by following the procedure outlined in \cite{PhysRevB.82.245417,PhysRevB.99.155149}.

Once again then, the \textit{weak damping} case $(\nu<\delta(\sim t^{-1}))$ yields for the current
\begin{equation}
I(t)=4eJe^{-\nu t}\sin\Big[\phi +\frac{\delta Jt}{\Delta}\Big]\Big[\cos(\rho t)+\frac{\nu}{\rho}\sin(\rho t)\Big],\label{stp12}
\end{equation}
where $\rho = \sqrt{\delta^{2}-\nu^{2}}$; while \textit{strong damping} $(\nu>>\delta)$ leads to
\begin{equation}
I(t) = 4eJ\sin\Big[\phi +\frac{\delta Jt}{\Delta}\Big]e^{-\frac{\delta^{2}}{2\nu}t}.\label{stp13}
\end{equation}

To facilitate numerical comparison between the Gaussian and the symmetric TP cases, we weigh the results in Eq. (\ref{gaussian6}) and (\ref{stp13}) respectively and set $\gamma=2\nu$ in Fig. \ref{current} for the entire range of damping.

We now compare the current for the Gaussian case with that for the symmetric TP in Fig. \ref{currentweak} for \textit{weak dissipation} and in Fig. \ref{currentstrong} for \textit{strong dissipation}. With the identification of the respective parameters as $\gamma$ and $\nu$ we find that the results have similarities as well as dissimilarities. For \textit{weak dissipation}, though the profiles of current are almost overlapping for very small period of initial time, they differ by a large margin with increasing time. The reason is clear: for TP in weak dissipation, relaxation occurs in the presence of the superposition of two distinct frequencies $\pm\delta$. On the other hand, for the case of \textit{strong dissipation}, both of the current profiles exactly overlap with each other as they follow the exact same equations (Eq. (\ref{gaussian6}) and (\ref{stp13})).
\subsection{Telegraphic process -- asymmetric case}
\label{ASTP}
In the more general case of asymmetric jumps, $p_{+} \neq p_{-}$, and we may introduce an asymmetry parameter $q = p_{+}-p_{-}$. From Eq. (\ref{stp08}) and (\ref{stp09}) we now have
\begin{equation}
[\boldsymbol{U}(s)]_{av} = \frac{(s+2\nu)+iq\delta}{s(s+2\nu)+\delta(\delta-i2\nu q)},\label{astp01}
\end{equation}
The Laplace-inversion of Eq. (\ref{astp01}) reads
\begin{align}
[\boldsymbol{U}(t)]_{av} = &\frac{1}{2}\Big\{1-\frac{\nu+iq\delta }{\sqrt{\nu^{2}-(\delta')^{2}}}\Big\}e^{-\nu t-\sqrt{\nu^{2}-(\delta')^{2}}t}\nonumber\\
 +&\frac{1}{2}\Big\{1+\frac{\nu+iq\delta }{\sqrt{\nu^{2}-(\delta')^{2}}}\Big\}e^{-\nu t+\sqrt{\nu^{2}-(\delta')^{2}}t},\label{astp02}
\end{align}
where $(\delta')^{2}=\delta(\delta-i2\nu q)=\delta^2 -i2\nu q\delta$.

Once again, the same expression can be directly arrived at by employing exponentiation of $2\times2$ matrices as shown in \cite{PhysRevB.82.245417,PhysRevB.99.155149}.

Now $[\boldsymbol{U}(t)]_{av}$ is complex and hence its real and imaginary parts have to be dealt with separately to yield for the current 
\begin{align}
I(t)= 4eJ\bigg\{&\sin\Big[\phi+\frac{\delta Jt}{\Delta}\Big]Re_{[\boldsymbol{U}(t)]_{av}}\nonumber\\
+&\cos\Big[\phi+\frac{\delta Jt}{\Delta}\Big]Im_{[\boldsymbol{U}(t)]_{av}}\bigg\},\label{astp03}
\end{align}

One can easily calculate the real and imaginary part of $[\boldsymbol{U}(t)]_{av}$, i.e., $Re_{[\boldsymbol{U}(t)]_{av}}$ and $Im_{[\boldsymbol{U}(t)]_{av}}$, starting from Eq. (\ref{astp02}).

\begin{figure*}[]
     \centering
     \subfloat[]{\includegraphics[width=0.52\textwidth]{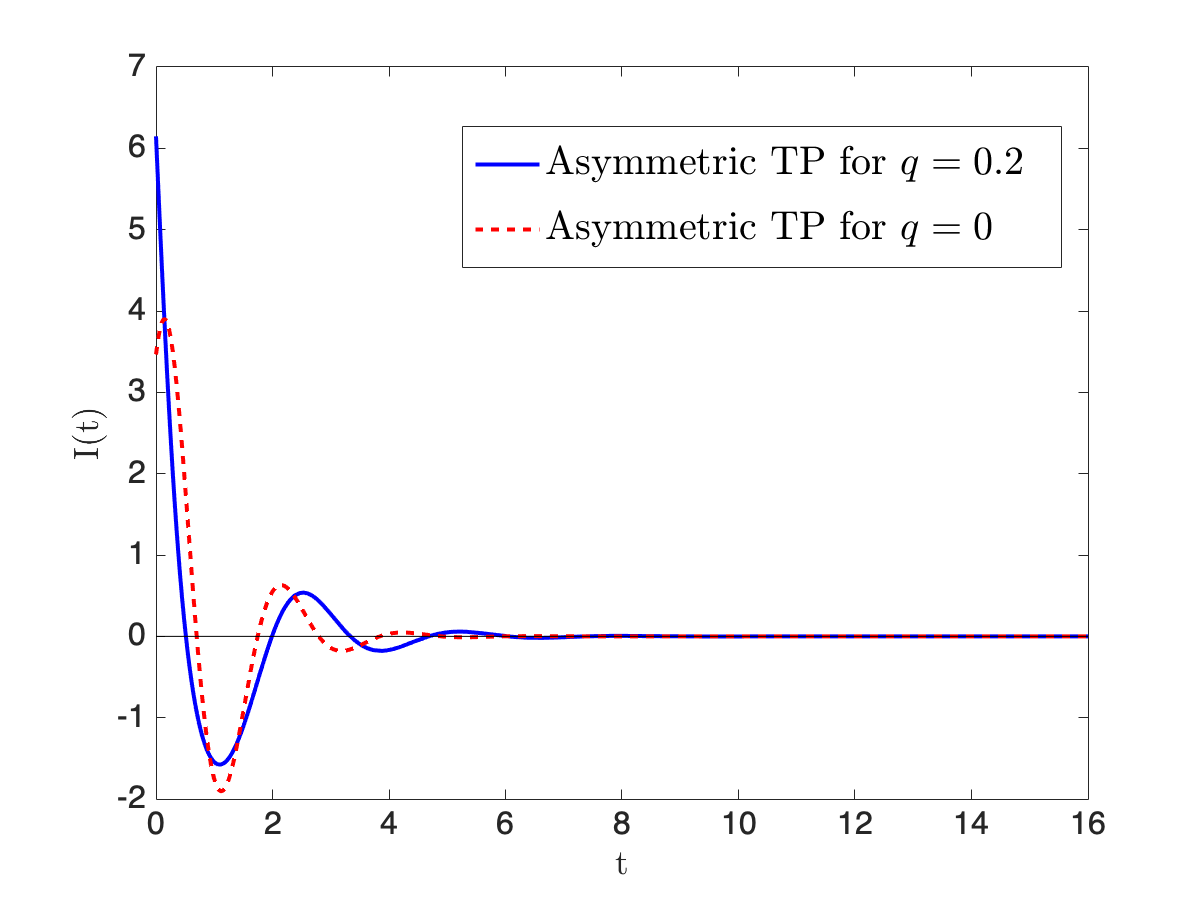}\label{currentweak2}}
     \subfloat[]{\includegraphics[width=0.52\textwidth]{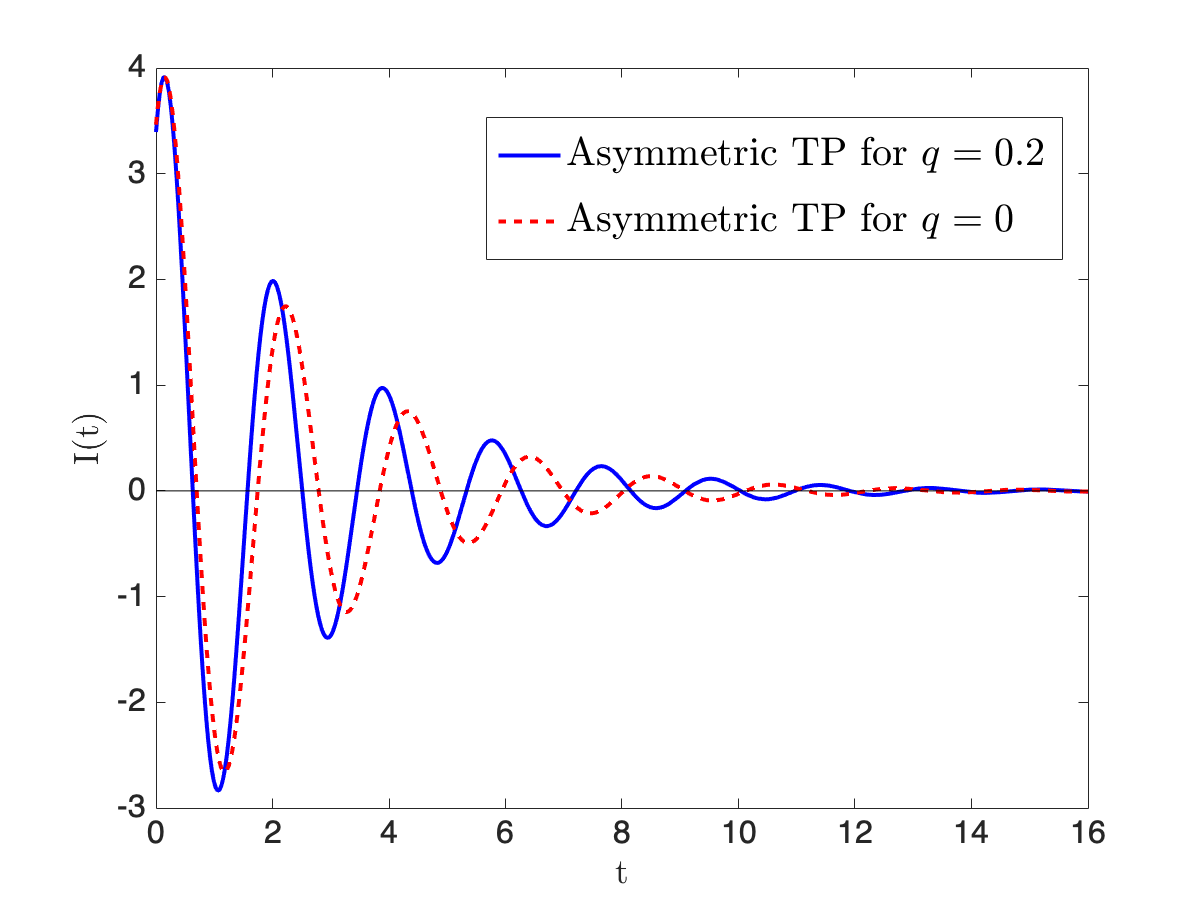}\label{currentstrong2}}
     \caption{(color online) Plot of Persistent current with time for Asymmetric TP with $q=0.2$ (solid blue line) and Asymmetric TP with $q=0$ (dashed red line): (a) for \textit{weak dissipation}, the dissipation parameter is $\nu=\frac{5}{4}$ (b) for \textit{strong dissipation}, the parameter is  $\nu= 3.$ For both of the plots, $e=1=J, \Delta = \frac{1}{2}, \phi = \frac{\pi}{3},$ and $\delta$ is given in Eq. (\ref{gaussian3}).}\label{current2}
\end{figure*}

At this point, we present the current for the case of asymmetric TP in Fig. \ref{currentweak2} for \textit{weak dissipation} and in Fig. \ref{currentstrong2} for \textit{strong dissipation}. We have plotted the current for the asymmetric TP for two different values of the asymmetry parameter $q$. The solid blue line is for asymmetric TP with $q = 0.2$ for both \textit{weak} and \textit{strong dissipation} limit. The dashed red line is for asymmetric TP with $q = 0$ (again, for both the limiting cases) which agrees with that of the symmetric TP when the \textit{a-priori} probabilities are the same. This fact is indeed true, as for $q=0$, Eq. (\ref{astp01}) and (\ref{astp02}) immediately reduce to Eq. (\ref{stp10}) and (\ref{stp11}) respectively. Also, by comparing Fig. \ref{currentweak2} and \ref{currentstrong2}, we can conclude that current relaxes at a higher value of time in the limiting case of \textit{strong dissipation} than that of the \textit{weak dissipation}. 
\section{Bond current under a quantum noise}
\label{quantum}
For describing decoherence properties in the realm of quantum mechanics it is important to view the environment as a quantum heat bath unlike the classical cases treated in Sec. \ref{ClassicalNoise}. We attempt to do that here by viewing the fluctuating term $f(t)$ as a two-state process, like the telegraph situation, but now given explicitly in terms of Pauli pseudo-spin operators which are coupled to bosonic fields as in the celebrated spin-boson model for dissipative quantum systems \cite{doi:10.1142/8334, DattaguptaPuri, RevModPhys.59.1}. The ring Hamiltonian is then a generalized form of Eq. (\ref{intro11}): 
\begin{align}
\boldsymbol{\mathcal{H}_{R}} = &-[J+\Delta \boldsymbol{\sigma}_{z}]\boldsymbol{\it{\Pi}} +\frac{E_{0}}{2}\boldsymbol{\sigma}_{z}+\boldsymbol{\sigma}_{x}\sum_{j}g_{j}(\boldsymbol{b}_{j}+\boldsymbol{b}_{j}^{\dagger})\nonumber\\
&+\sum_{j}\omega_{j}\boldsymbol{b}_{j}^{\dagger}\boldsymbol{b}_{j}.\label{quan01}
\end{align}
The pseudo spin operator $\boldsymbol{\sigma}_{z}$ (with the pre-factor $\Delta$) plays the role of $f(t)$ in the TP wherein the time dependence of the latter is viewed to arise from the interaction picture representation:
\begin{equation}
\boldsymbol{\sigma}_{z}(t) = e^{it\big(\frac{E_{0}}{2}\boldsymbol{\sigma}_{z}+\boldsymbol{\mathcal{H}_{I}}+\boldsymbol{\mathcal{H}_{B}}\big)}\boldsymbol{\sigma}_{z}(0)e^{-it\big(\frac{E_{0}}{2}\boldsymbol{\sigma}_{z}+\boldsymbol{\mathcal{H}_{I}}+\boldsymbol{\mathcal{H}_{B}}\big)},\label{quan02}
\end{equation}
where $\boldsymbol{\mathcal{H}_{I}}$ and $\boldsymbol{\mathcal{H}_{B}}$ are the respective coupling Hamiltonian and the bath Hamiltonian, given by
\begin{equation}
\boldsymbol{\mathcal{H}_{I}} = \boldsymbol{\sigma}_{x}\sum_{j}g_{j}(\boldsymbol{b}_{j}+\boldsymbol{b}_{j}^{\dagger}),~~ \boldsymbol{\mathcal{H}_{B}} = \sum_{j}\omega_{j}\boldsymbol{b}_{j}^{\dagger}\boldsymbol{b}_{j},\label{quan03}
\end{equation}
$g_{i}$'s being the coupling constants and $\boldsymbol{b}_{j}(\boldsymbol{b}_{j}^{\dagger})$ are bosonic operators driven by a harmonic bath characterized by frequencies $\omega_{j}$. Thus, the expanded ring Hamiltonian now lives in an extended, product Hilbert space of $\{\boldsymbol{\it{\Pi}}\otimes\boldsymbol{\sigma}\otimes\boldsymbol{b}\}$. The underlying idea is that the operator $\boldsymbol{\sigma}_{x}$, being purely off-diagonal in the representation of $\boldsymbol{\sigma}_{z}$, causes transitions of the latter between $+1$ and $-1$ and vice-versa much akin to the telegraph process and these transitions are in turn influenced by fluctuating bosonic fields $\boldsymbol{b}_{j}(\boldsymbol{b}_{j}^{\dagger})$ triggered by the bath $\boldsymbol{\mathcal{H}_{B}}$. The transitions are in general asymmetric because of the presence of the energy term $E_{0}$ which can be comparable to the thermal energy, leading to detailed-balance factors(cf., Eq. (\ref{stp05})), as argued in \cite{2017,2019}. Our analysis here is very much in the spirit of a corresponding treatment for a qubit \cite{kamilSDG}, though we prefer to work here in the Heisenberg picture as opposed to the Schrödinger picture, and more importantly, treat the term proportional to $\Delta$ exactly in order to make contact with the TP, especially for weak damping.

As indicated in Eq. (\ref{intro13}) we would want to calculate the time-dependence of the current operator via  
\begin{equation}
\boldsymbol{j}_{12}(t) = \exp{(i\boldsymbol{\mathcal{H}_{R}}t)}\boldsymbol{j}_{12}(0)\exp{(-i\boldsymbol{\mathcal{H}_{R}}t)},\label{quan04}
\end{equation}
and perform a statistical mechanical average in the canonical ensemble with the aid of an equilibrium density operator $\boldsymbol{\rho}_{B}$ defined below. Because $\boldsymbol{\mathcal{H}_{R}}$ (cf., Eq. (\ref{quan01})) contains a set of non-commuting operators the two exponentiated operators to the left and to the right of $\boldsymbol{j}_{12}$ will have to be averaged over in juxtaposition which is a complicated task. Instead, therefore, we would write the time-development in Eq. (\ref{quan04}) in terms of a single exponentiated operator with the aid of a Liouvillian $\boldsymbol{\mathcal{H}_{R}}^{\times}$ associated with $\boldsymbol{\mathcal{H}_{R}}$ though the price we pay is that we are led to operate within a larger Hilbert space\cite{doi:10.1063/1.1703941,dattagupta1987relaxation}.

Retracing the steps as in Eq. (\ref{intro16}) and employing the first-principles Hamiltonian as in Eq. (\ref{quan01}) above, we may write in the Liouvillian formalism (cf., Eq. $(7.19)$ of \cite{DattaguptaPuri})
\begin{align}
&\boldsymbol{\langle j}_{12}(t)\boldsymbol{\rangle} =\nonumber\\
&\exp{\Big(itJ\frac{\delta}{\Delta}\Big)}\sum_{\mu\mu'}\big(+\mu,-\mu\big|\exp{[t(i\boldsymbol{\mathcal{L}_{s}}-\boldsymbol{\Sigma})]}\big|+\mu',-\mu'\big),\label{quan05}
\end{align}
where, we recall that $\ket{+}$ and $\ket{-}$ are the eigenstates of $\boldsymbol{\it{\Pi}}$. The Greek indices $\mu,\mu'$, etc., are used to denote the eigenstates of $\boldsymbol{\sigma}_{z}$, i.e., 
\begin{equation}
\boldsymbol{\sigma}_{z}\ket{\uparrow} = \ket{\uparrow},~~\boldsymbol{\sigma}_{z}\ket{\downarrow} = -\ket{\downarrow},\label{quan06}
\end{equation}
 Here (and below, the script $\boldsymbol{\mathcal{L}}$'s represent the Liouvillians associated with the corresponding ordinary operators)
 \begin{equation}
 \boldsymbol{\mathcal{L}_{s}} = -\Delta(\boldsymbol{\sigma}_{z}\boldsymbol{\it{\Pi}})^{\times}+\bigg(\frac{E_{0}}{2}\bigg)\boldsymbol{\sigma}_{z}^{\times},\label{quan07}
 \end{equation}
the superscript $\times$ denoting the corresponding Liouville operator and $\boldsymbol{\Sigma}$ the Markovian limit of the self-energy operator (cf., Eq. $(7.20)$ of \cite{DattaguptaPuri}): 
\begin{equation}
\boldsymbol{\Sigma} = \sum_{bb'}\big(b,b\big|\int_{0}^{\infty}dt\{\boldsymbol{\mathcal{L}_{I}}e^{[i(\boldsymbol{\mathcal{L}_{s}}+\boldsymbol{\mathcal{L}_{B}})t]}\boldsymbol{\mathcal{L}_{I}}\}\big|b',b'\big)\langle b|\boldsymbol{\rho}_{B}|b\rangle,\label{quan08}
\end{equation}
$\boldsymbol{\rho}_{B}$ being the density matrix for the bath:
\begin{equation}
\boldsymbol{\rho}_{B} = \frac{\exp{(-\frac{\boldsymbol{\mathcal{H}_{B}}}{k_{B}T})}}{Tr[\exp{(-\frac{\boldsymbol{\mathcal{H}_{B}}}{k_{B}T})}]},\label{quan09}
\end{equation}
corresponding to the temperature $T$ of the bath and $k_{B}$ is the \textit{Boltzmann} constant here.

Once again, the $\boldsymbol{\mathcal{L}}$'s represent Liouville operators -- $\boldsymbol{\mathcal{L}_{I}}$  is associated with $\boldsymbol{\mathcal{H}_{I}}$ while $\boldsymbol{\mathcal{L}_{B}}$ is associated with $\boldsymbol{\mathcal{H}_{B}}$.

The formalism outlined here in which $\boldsymbol{\mathcal{L}_{I}}$ is treated to second order in the Born approximation, is in the spirit of Lindblad, among others \cite{Lindblad}. As argued in Chap. $8.4$ of \cite{DattaguptaPuri}, the only relevant matrix elements in Eq. (\ref{quan05}) are
\begin{align}
&\big(+\uparrow,-\uparrow\big|...\big|+\uparrow,-\uparrow\big), \big(+\uparrow,-\uparrow\big|...\big|+\downarrow,-\downarrow\big)\nonumber\\
& \big(+\downarrow,-\downarrow\big|...\big|+\uparrow,-\uparrow\big),~\textnormal{and}~\big(+\downarrow,-\downarrow\big|...\big|+\downarrow,-\downarrow\big).\label{quan10}
\end{align}
We find (cf., Appendix $7.A$ of \cite{DattaguptaPuri})
\begin{align}
&\big(+\uparrow,-\uparrow\big|(i\boldsymbol{\mathcal{L}_{s}}-\boldsymbol{\Sigma})\big|+\uparrow,-\uparrow\big)= (-i\delta+2\nu p_{-}),\nonumber\\
& \big(+\uparrow,-\uparrow\big|(i\boldsymbol{\mathcal{L}_{s}}-\boldsymbol{\Sigma})\big|+\downarrow,-\downarrow\big)= -2\nu p_{-},\nonumber\\
& \big(+\downarrow,-\downarrow\big|(i\boldsymbol{\mathcal{L}_{s}}-\boldsymbol{\Sigma})\big|+\uparrow,-\uparrow\big)= -2\nu p_{+}, ~\textnormal{and}\nonumber\\
&\big(+\downarrow,-\downarrow\big|(i\boldsymbol{\mathcal{L}_{s}}-\boldsymbol{\Sigma})\big|+\downarrow,-\downarrow\big)= (i\delta+2\nu p_{+}),\label{quan11}
\end{align}
where $\delta = 2\sqrt{3}\Delta \sin{\phi}$, and the average probabilities are given by the normalized Boltzmann factors \cite{PhysRevB.49.13728, PhysRevB.52.12126}
\begin{equation}
p_{\pm} = \frac{\exp\big(\pm\frac{E_{0}}{2k_{B}T}\big)}{2\cosh\big(\frac{E_{0}}{2k_{B}T}\big)}.\label{quan12}
\end{equation}
Reverting to Sec. \ref{ASTP} it is clear that the right-hand side of Eq. (\ref{quan05}) is identical in structure to the $2\times2$ matrix form of $[\boldsymbol{U}(t)]_{av}$, if we identify $-\boldsymbol{\Sigma}$ with the rate matrix $\boldsymbol{W}$ and hence, the rest of the analysis for the current is the same as for the case of the TP. A similar connection between decoherence and relaxation of a qubit (in contact with the environment of a single electron transistor) and the spectral density of a spin-boson model in the weak-coupling limit, was shown earlier in \cite{PhysRevB.77.075325}. The redeeming feature of the present treatment however is that the underlying relaxation rate $\nu$ has got a microscopic structure in terms of the parameters of the bosonic bath, such that in the weak-coupling and the Markovian limit (cf., Eq. $(1.94)$ and $(8.71)$ of \cite{DattaguptaPuri})
\begin{equation}
2\nu = \sum_{j}g_{j}^{2}\int_{-\infty}^{\infty}d\tau[\boldsymbol{\langle b}_{j}(0)\boldsymbol{b}_{j}^{\dagger}(t)\boldsymbol{\rangle}_{B}+\boldsymbol{\langle b}_{j}^{\dagger}(0)\boldsymbol{b}_{j}(t)\boldsymbol{\rangle}_{B}],\label{quan13}
\end{equation}
where the subscript $B$ denotes thermal average over the bosonic bath.
\section{Summary and concluding remarks}
\label{conclusions}
The fascinating effect of the Aharonov-Bohm phase on the bond current in a three-site mesoscopic ring is first exactly studied. The bond current being a quantum coherent property is expected to undergo decoherence and relaxation when the ring is put in contact with a heat bath. The resultant decay of the current is first studied when the environment is viewed to cause either Gaussian modulation or discrete jump modulation (under a telegraph process) within a stochastic modeling of the bath. Comparative figures are employed to distinguish between the Gaussian and telegraph cases both in the \textit{weak damping} and in the \textit{strong damping} limits. Having derived these results analytically, we turn our attention next to the situation in which the heat bath has to be treated quantum mechanically, in order to investigate decoherence effects in the quantum domain, which are expected to be significant at low temperatures. The much-studied spin-boson model of a dissipative quantum system is found to be eminently suitable for critically assessing the limits of validity of the classical telegraph process. While we do not find it relevant here to delve into the detailed structure of the relaxation rate such an analysis can be easily carried out on the basis of whether the relaxation is driven by electron or by phonon processes, within the framework of the spin-boson model (as in \cite{DattaguptaPuri}). Our derived results for the time-dependence of the bond current are expected to be of interest in the contemporary topic of coherence-to-decoherence transition in mesoscopic devices. 
\section*{Acknowledgment}
SD is grateful to the Indian National Science Academy for support through their Senior Scientist scheme. He would also like to thank Ora Entin-Wohlman, Shmuel Gurvitz and Amnon Aharony for introducing him to the topic and for insightful discussions.


%

\end{document}